\documentstyle[12pt]{article}

\catcode`@=12
\begin{document}
\thispagestyle{empty}
\begin{flushright} UCRHEP-T182\\December 1996\
\end{flushright}
\vspace{0.5in}
\begin{center}
{\Large \bf Supersymmetric Scalar Masses, Z', and E(6)\\}
\vspace{1.0in}
{\bf Ernest Ma\\}
\vspace{0.3in}
{\sl Department of Physics, University of California\\}
{\sl Riverside, CA 92521, USA\\}
\vspace{1.0in}
\end{center}
\begin{abstract}\
Assuming the existence of a supersymmetric U(1) gauge factor at the TeV 
energy scale (motivated either by the superstring-inspired $E_6$ model 
or low-energy electroweak phenomenology), several important consequences 
are presented.  The two-doublet Higgs structure at the 100 GeV energy scale 
is shown to be different from that of the Minimal Supersymmetric Standard 
Model (MSSM).  A new neutral gauge boson $Z'$ corresponding to the extra 
U(1) mixes with the $Z$.  The supersymmetric scalar quarks and leptons 
receive new contributions to their masses from the spontaneous breaking 
of this extra U(1).  The assumption of universal soft supersymmetry 
breaking terms at the grand-unification energy scale implies a connection 
between the U(1) breaking scale and the ratio of the vacuum expectation 
values of the two electroweak Higgs doublets.
\end{abstract}
\vspace{0.3in}
\noindent {\it To appear in Proc. of the First Latin American Symposium on 
High Energy Physics, Merida, Mexico (November, 1996)}
\newpage
\baselineskip 24pt
\section*{Introduction}

Consider the sequential reduction in rank of the symmmetry group $E_6$:
\begin{eqnarray}
E_6 &\rightarrow& SO(10) ~[\times U(1)_\psi] \\ 
    &\rightarrow& SU(5) ~[\times U(1)_\chi] \\
    &\rightarrow& SU(3)_C \times SU(2)_L ~[\times U(1)_Y].
\end{eqnarray}
At each step, a U(1) gauge factor may or may not appear, depending on the 
details of the symmetry breaking.  If $E_6$ is indeed the grand-unification 
group, it is often assumed that a single U(1) survives\cite{1} down to the 
TeV energy range, given by
\begin{equation}
U(1)_\psi \times U(1)_\chi \rightarrow U(1)_\alpha.
\end{equation}
This talk is concerned mainly with the phenomenological consequences of 
extending the MSSM to include this $U(1)_\alpha$.

\section*{New U(1) and New Particles}

Under the maximal subgroup $SU(3)_C \times SU(3)_L \times SU(3)_R$, the 
fundamental representation of $E_6$ is given by
\begin{equation}
{\bf 27} = (3, 3, 1) + (3^*, 1, 3^*) + (1, 3^*, 3).
\end{equation}
Under the subgroup $SU(5) \times U(1)_\psi \times U(1)_\chi$, we then have 
\begin{eqnarray}
{\bf 27} &=& (10; 1, -1) ~[(u,d),u^c,e^c] \nonumber \\
         &+& (5^*; 1, 3) ~[d^c,(\nu_e,e)] \nonumber \\
         &+& (1; 1, -5) ~[N] \nonumber \\
         &+& (5; -2, 2) ~[h,(E^c,N_E^c)] \nonumber \\
         &+& (5^*; -2, -2) ~[h^c,(\nu_E,E)] \nonumber \\
         &+& (1; 4, 0) ~[S],
\end{eqnarray}
where the U(1) charges refer to $2 \sqrt 6 Q_\psi$ and $2 \sqrt {10} Q_\chi$. 
Note that the known quarks and leptons are contained in $(10; 1, -1)$ and 
$(5^*; 1, 3)$, and the two Higgs scalar doublets are represented by 
$(\nu_E,E)$ and $(E^c,N_E^c)$.  Let
\begin{equation}
Q_\alpha = Q_\psi \cos \alpha - Q_\chi \sin \alpha,
\end{equation}
then the so-called $\eta$-model\cite{1,2} is obtained with $\tan \alpha = 
\sqrt {3/5}$ and we have
\begin{eqnarray}
{\bf 27} &=& (10; 2) + (5^*; -1) + (1; 5) \nonumber \\
         &+& (5; -4) + (5^*; -1) + (1; 5),
\end{eqnarray}
where $2 \sqrt {15} Q_\eta$ is denoted; 
and the $N$-model\cite{3} is obtained with $\tan \alpha = -1/\sqrt{15}$ 
resulting in
\begin{eqnarray}
{\bf 27} &=& (10; 1) + (5^*; 2) + (1; 0) \nonumber \\
         &+& (5; -2) + (5^*; -3) + (1; 5),
\end{eqnarray}
where $2 \sqrt{10} Q_N$ is denoted.  The $\eta$-model is theoretically 
attractive because it is obtained if the symmetry breaking of $E_6$ 
occurs via only the adjoint {\bf 78} representation which is what the 
superstring flux mechanism may do\cite{1}.  It is also phenomenologically 
interesting because it allows for an explanation of the experimental 
$R_b$ excess\cite{2}. 
The $N$-model is so called because $N$ has $Q_N = 0$.  It allows 
$S$ to be a naturally light singlet neutrino and is ideally suited to explain 
the totality of all neutrino-oscillation experiments\cite{3}.  It is also
a natural consequence of an alternative $SO(10)$ decomposition\cite{4} of 
$E_6$, {\it i.e.}
\begin{eqnarray}
{\bf 16} &=& [(u,d), u^c, e^c; h^c, (\nu_E, E); S], \\
{\bf 10} &=& [h, (E^c, N_E^c); d^c, (\nu_e, e)], \\
 {\bf 1} &=& [N],
\end{eqnarray}
which differs from the conventional assignment by how the $SU(5)$ 
multiplets are embedded.

\section*{Higgs Sector}

The Higgs sector of the $U(1)_\alpha$-extended supersymmetric model 
consists of two doublets and a singlet.  They transform under 
$SU(3)_C \times SU(2)_L \times U(1)_Y \times U(1)_\alpha$ as follows.
\begin{equation}
\tilde \Phi_1 \equiv \left( \begin{array} {c} \bar \phi_1^0 \\ -\phi_1^- 
\end{array} \right) \equiv \left( \begin{array} {c} \tilde \nu_E \\ \tilde E 
\end{array} \right) \sim \left( 1, 2, -{1 \over 2}; -{1 \over \sqrt 6} 
\cos \alpha + {1 \over \sqrt {10}} \sin \alpha \right),
\end{equation}
\begin{equation}
\Phi_2 \equiv \left( \begin{array} {c} \phi_2^+ \\ \phi_2^0 \end{array} 
\right) \equiv \left( \begin{array} {c} \tilde E^c \\ \tilde N_E^c \end{array} 
\right) \sim \left( 1, 2, {1 \over 2}; -{1 \over \sqrt 6} \cos \alpha - 
{1 \over \sqrt {10}} \sin \alpha \right),
\end{equation}
\begin{equation}
\chi \equiv \tilde S \sim \left( 1, 1, 0; \sqrt {2 \over 3} \cos \alpha 
\right).
\end{equation}
Hence the Higgs potential has the contribution
\begin{equation}
V_F = f^2 [(\Phi_1^\dagger \Phi_2)(\Phi_2^\dagger \Phi_1) + (\Phi_1^\dagger 
\Phi_1 + \Phi_2^\dagger \Phi_2)(\bar \chi \chi)],
\end{equation}
where $f$ is the Yukawa coupling of the $\tilde \Phi_1 \Phi_2 \chi$ term 
in the superpotential.  From the gauge interactions, we have the additional 
contribution
\begin{eqnarray}
V_D &=& {1 \over 8} g_2^2 [(\Phi_1^\dagger \Phi_1)^2 + (\Phi_2^\dagger 
\Phi_2)^2 + 2 (\Phi_1^\dagger \Phi_1)(\Phi_2^\dagger \Phi_2) - 4 
(\Phi_1^\dagger \Phi_2)(\Phi_2^\dagger \Phi_1)] \nonumber \\ &+& {1 \over 2} 
g_1^2 [ -{1 \over 2} \Phi_1^\dagger \Phi_1 + {1 \over 2} \Phi_2^\dagger \Phi_2 
]^2 + {1 \over 2} g_\alpha^2 [ \left( -{1 \over \sqrt 6} \cos \alpha + 
{1 \over \sqrt {10}} \sin \alpha \right) \Phi_1^\dagger \Phi_1 \nonumber \\ 
&~& + \left( -{1 \over \sqrt 6} 
\cos \alpha - {1 \over \sqrt {10}} \sin \alpha \right) \Phi_2^\dagger \Phi_2 
+ \sqrt {2 \over 3} \cos \alpha ~\bar \chi \chi]^2.
\end{eqnarray}
Let $\langle \chi \rangle = u$, then $\sqrt 2 Re \chi$ is a physical scalar 
boson with
\begin{equation}
M^2 = {4 \over 3} \cos^2 \alpha ~g_\alpha^2 u^2,
\end{equation}
and the $(\Phi_1^\dagger \Phi_1) \sqrt 2 Re \chi$ coupling is
\begin{equation}
F = \sqrt 2 u \left[ f^2 + g_\alpha^2 \sqrt {2 \over 3} \cos \alpha \left( 
-{1 \over \sqrt 6} \cos \alpha + { 1 \over \sqrt {10}} \sin \alpha \right) 
\right].
\end{equation}
The effective $(\Phi_1^\dagger \Phi_1)^2$ coupling $\lambda_1$ is thus 
given by\cite{5,6,7}
\begin{eqnarray}
\lambda_1 &=& {1 \over 4} (g_1^2 + g_2^2) + g_\alpha^2 \left( -{1 \over 
\sqrt 6} \cos \alpha + {1 \over \sqrt {10}} \sin \alpha \right)^2 - {F^2 
\over M^2} \nonumber \\ &=& {1 \over 4} (g_1^2 + g_2^2) + \left( 1 - \sqrt 
{3 \over 5} \tan \alpha \right) f^2 - {{3 f^4} \over {2 \cos^2 \alpha 
~g_\alpha^2}}.
\end{eqnarray}
Similarly,
\begin{eqnarray}
\lambda_2 &=& {1 \over 4} (g_1^2 + g_2^2) + \left( 1 + \sqrt {3 \over 5} 
\tan \alpha \right) f^2 - {{3 f^4} \over {2 \cos^2 \alpha ~g_\alpha^2}}, \\ 
\lambda_3 &=& -{1 \over 4} g_1^2 + {1 \over 4} g_2^2 + f^2 - {{3 f^4} \over 
{2 \cos^2 \alpha ~g_\alpha^2}}, \\ \lambda_4 &=& -{1 \over 2} g_2^2 + f^2,
\end{eqnarray}
where the effective two-doublet Higgs potential has the generic form
\begin{eqnarray}
V &=& m_1^2 \Phi_1^\dagger \Phi_1 + m_2^2 \Phi_2^\dagger \Phi_2 + m_{12}^2 
(\Phi_1^\dagger \Phi_2 + \Phi_2^\dagger \Phi_1) \nonumber \\ &+& {1 \over 2} 
\lambda_1 (\Phi_1^\dagger \Phi_1)^2 + {1 \over 2} (\Phi_2^\dagger \Phi_2)^2 
+ \lambda_3 (\Phi_1^\dagger \Phi_1)(\Phi_2^\dagger \Phi_2) + \lambda_4 
(\Phi_1^\dagger \Phi_2)(\Phi_2^\dagger \Phi_1).
\end{eqnarray}
From Eqs.~(20) to (23), it is clear that the MSSM is recovered in the limit 
of $f = 0$. 
Let $\langle \phi^0_{1,2} \rangle \equiv v_{1,2}$, $\tan \beta \equiv v_2/
v_1$, and $v^2 \equiv v_1^2 + v_2^2$, then this $V$ has an upper bound on 
the lighter of the two neutral scalar bosons given by
\begin{equation}
(m_h^2)_{max} = 2 v^2 [\lambda_1 \cos^4 \beta + \lambda_2 \sin^4 \beta + 
2 (\lambda_3 + \lambda_4) \sin^2 \beta \cos^2 \beta] + \epsilon,
\end{equation}
where we have added the radiative correction due to the $t$ quark and 
its supersymmetric scalar partners, {\it i.e.}
\begin{equation}
\epsilon = {{3 g_2^2 m_t^4} \over {8 \pi^2 M_W^2}} \ln \left( 1 + {\tilde m^2 
\over m_t^2} \right).
\end{equation}
Using Eqs.~(20) to (23), we obtain
\begin{eqnarray}
&~& (m_h^2)_{max} = M_Z^2 \cos^2 2 \beta + \epsilon \nonumber \\ &+& 
{1 \over {\sqrt 2 G_F}} \left[ f^2 \left( {3 \over 2} - \sqrt {3 \over 5} 
\tan \alpha \cos 2 \beta - {1 \over 2} \cos^2 2 \beta \right) - {{3 f^4} 
\over {2 \cos^2 \alpha ~g_\alpha^2}} \right].
\end{eqnarray}
Hence the MSSM bound can be exceeded for a wide range of values of $\alpha$ 
and $\beta$.  Normalizing $U(1)_Y$ and $U(1)_\alpha$ at the grand-unification 
energy scale, we find it to be a very good approximation\cite{8} to have 
$g_\alpha^2 = (5/3) g_1^2$.  We use this and vary $f^2$ in Eq.~(27) subject 
to the condition that $V$ be bounded from below.  We find the largest 
numerical value of $m_h$ to be about 142 GeV, as compared to 128 GeV in 
the MSSM, and this is achieved with 
\begin{equation}
\tan \alpha = - {{2 \sqrt {3/5} \cos 2 \beta} \over {3 - \cos^2 2 \beta}},
\end{equation}
which is possible in the $\eta$-model.

\section*{Z - Z' Sector}

The new $Z'$ of this model mixes with the standard $Z$ so that the 
experimentally observed $Z$ is actually
\begin{equation}
Z_1 = Z \cos \theta + Z' \sin \theta,
\end{equation}
where
\begin{equation}
\theta \simeq -{1 \over 2} \sqrt {3 \over 2} {1 \over {\cos \alpha}} 
{g_Z \over g_\alpha} \left( \sin^2 \beta - {1 \over 2} + {1 \over 2} 
\sqrt {3 \over 5} \tan \alpha \right) {v^2 \over u^2},
\end{equation}
resulting in a slight shift of its mass from that predicted by the 
standard model, as well as a slight change in its couplings to the 
usual quarks and leptons.  These deviations can be formulated in terms 
of the oblique parameters\cite{6}:
\begin{eqnarray}
\epsilon_1 &=& \left[ \sin^4 \beta - {1 \over 4} \left( 1 - \sqrt {3 \over 5} 
\tan \alpha \right)^2 \right] {v^2 \over u^2} \simeq \alpha T, \\ 
\epsilon_2 &=& {1 \over 4} (3 - \sqrt {15} \tan \alpha) \left[ \sin^2 \beta 
- {1 \over 2} \left( 1 - \sqrt {3 \over 5} \tan \alpha \right) \right] 
{v^2 \over u^2} \simeq - {{\alpha U} \over {4 \sin^2 \theta_W}}, \\ 
\epsilon_3 &=& {1 \over 4} \left[ 1 - 3 \sqrt {3 \over 5} \tan \alpha + 
{1 \over {2 \sin^2 \theta_W}} \left( 1 + \sqrt {3 \over 5} \tan \alpha \right) 
\right] \nonumber \\ &\times& \left[ \sin^2 \beta - {1 \over 2} \left( 1 - 
\sqrt {3 \over 5} \tan \alpha \right) \right] {v^2 \over u^2} \simeq 
{{\alpha S} \over {4 \sin^2 \theta_W}}.
\end{eqnarray}
Note that for $\sin^2 \beta$ near $(1/2)(1-\sqrt {3/5} \tan \alpha)$, 
$\epsilon_{1,2,3}$ are all suppressed.  In any case, the experimental 
errors on these quantities are fractions of a percent, hence 
$u \sim$ TeV is allowed.

The mass of $Z'$ is approximately equal 
to that of $\sqrt 2 Re \chi$, {\it i.e.} $M$ of Eq.~(18).  Its interactions 
are of course determined by $U(1)_\alpha$.  In particular, in the $N$-model, 
two $S$'s are light singlet neutrinos, hence the ratio of the decay rates 
of $Z'$ to $\nu \bar \nu + S \bar S$ over $Z'$ to $\ell^- \ell^+$ 
is 62/15, instead of 4/5 without the $S$'s.  This would be a great 
experimental signature.

\section*{Supersymmetric Scalar Masses}

Consider the masses of the supersymmetric scalar partners of the quarks and 
leptons:
\begin{equation}
m_B^2 = m_0^2 + m_R^2 + m_F^2 + m_D^2,
\end{equation}
where $m_0$ is a universal soft supersymmetry breaking mass at the 
grand-unification scale, $m_R^2$ is a correction generated by the 
renormalization-group equations running from the grand-unification 
scale down to the TeV scale, $m_F$ is the explicit mass of the fermion 
partner, and $m_D^2$ is a term induced by gauge-boson masses.  In the MSSM, 
$m_D^2$ is of order $M_Z^2$ and does not change $m_B$ significantly.  In the 
$U(1)_\alpha$-extended model, $m_D^2$ is of order $M_{Z'}^2$ and will affect 
$m_B$ in a nontrivial way.  For example, for the ordinary quarks and leptons, 
\begin{eqnarray}
\Delta m_D^2 (10; 1, -1) &=& {1 \over 8} M_{Z'}^2 \left( 1 + \sqrt {3 \over 5} 
\tan \alpha \right), \\ \Delta m_D^2 (5^*; 1, 3) &=& {1 \over 8} M_{Z'}^2 
\left( 1 - 3 \sqrt {3 \over 5} \tan \alpha \right).
\end{eqnarray}
This would have important consequences on the experimental search of 
supersymmetric particles.  In fact, depending on $m_F$, it is possible 
for exotic scalars to be lighter than the usual scalar quarks and leptons.

Another important outcome of Eq.~(34) is that the $U(1)_\alpha$ and 
electroweak symmetry breakings are related\cite{9}.  To see this, go back to 
the two-doublet Higgs potential $V$ of Eq.~(24).  Using Eqs.~(20) to (23), we 
can express the parameters $m_{12}^2$, $m_1^2$, and $m_2^2$ in terms of the 
mass of the pseudoscalar boson, $m_A$, and $\tan \beta$.
\begin{eqnarray}
m_{12}^2 &=& -m_A^2 \sin \beta \cos \beta, \\ m_1^2 &=& m_A^2 \sin^2 \beta 
- {1 \over 2} M_Z^2 \cos 2 \beta \nonumber \\ &~& - {{2 f^2} \over g_Z^2} 
M_Z^2 \left[ 2 \sin^2 \beta + \left( 1 - \sqrt {3 \over 5} \tan \alpha \right) 
\cos^2 \beta - {{3 f^2} \over {2 \cos^2 \alpha ~g_\alpha^2}} \right], \\ 
m_2^2 &=& m_A^2 \cos^2 \beta + {1 \over 2} M_Z^2 \cos 2 \beta \nonumber \\ 
&~& - {{2 f^2} \over g_Z^2} M_Z^2 \left[ 2 \cos^2 \beta + \left( 1 + 
\sqrt {3 \over 5} \tan \alpha \right) \sin^2 \beta - {{3 f^2} \over {2 
\cos^2 \alpha ~g_\alpha^2}} \right].
\end{eqnarray}
On the other hand, using Eq.~(34), we have
\begin{eqnarray}
m_{12}^2 &=& f A_f u, \\ m_1^2 &=& m_0^2 + m_R^2 (\tilde g, f) + f^2 u^2 - 
{1 \over 4} \left( 1 - \sqrt {3 \over 5} \tan \alpha \right) M_{Z'}^2, \\ 
m_2^2 &=& m_0^2 + m_R^2 (\tilde g, f) + f^2 u^2 - {1 \over 4} \left( 1 + 
\sqrt {3 \over 5} \tan \alpha \right) M_{Z'}^2 + m_R^2 (\lambda_t),
\end{eqnarray}
where $f A_f$ is the coupling of the soft supersymmetry 
breaking $\tilde \Phi_1 \Phi_2 \chi$ scalar term, $\tilde g$ is 
the gluino, and $\lambda_t$ is the Yukawa coupling of $\Phi_2$ to the $t$ 
quark.  Matching Eqs.~(37) to (39) with Eqs.~(40) to (42) allows us to 
determine $u$ and $\tan \beta$ as a function of $f$ for a given set of 
parameters at the grand-unification scale.

In the MSSM assuming Eq.~(34),
\begin{equation}
m_1^2 - m_2^2 = - m_R^2 (\lambda_t) = - (m_A^2 + M_Z^2) \cos 2 \beta.
\end{equation}
Since $m_R^2 (\lambda_t) < 0$, we must have $\tan \beta > 1$.  In the 
$U(1)_\alpha$-extended model, because of the extra D-term contribution, 
$\tan \beta < 1$ becomes possible.  Another consequence is that because 
of Eq.~(35), a light scalar $t$ quark is not possible unless $\tan \alpha 
< -\sqrt {5/3}$.

\section*{Conclusions}

(1) Supersymmetric $U(1)_\alpha$ from $E_6$ is a good possiblity at the 
TeV scale. (2) The two-doublet Higgs structure at around 100 GeV will be 
different from that of the MSSM. (3) Supersymmetric scalar masses depend 
crucially on $U(1)_\alpha$. (4) The $U(1)_\alpha$ breaking scale and 
$\tan \beta$ are closely related.

\section*{Acknowledgments}

I thank Juan Carlos D'Olivo, Miguel 
Perez, and Rodrigo Huerta for their great hospitality and a stimulating 
symposium.  This work was supported in part by the U. S. Department of 
Energy under Grant No. DE-FG03-94ER40837.

\newpage
\bibliographystyle{unsrt}

\end{document}